\documentclass[10pt,conference]{IEEEtran}
\IEEEoverridecommandlockouts

\def\BibTeX{{\rm B\kern-.05em{\sc i\kern-.025em b}\kern-.08em
    T\kern-.1667em\lower.7ex\hbox{E}\kern-.125emX}}
    
\usepackage{cite}
\usepackage{amsmath,amssymb,amsfonts}
\usepackage{algorithmic}
\usepackage{graphicx}
\usepackage{textcomp}
\usepackage{tcolorbox}
\usepackage{booktabs}
\usepackage{colortbl}
\usepackage{graphicx}
\usepackage{xcolor}
\usepackage{enumitem}

\usepackage{times}
\usepackage{latexsym}
\usepackage{listings}
\usepackage{xspace}
\usepackage{tabularx}
\usepackage{comment}
\usepackage{setspace}
\usepackage[T1]{fontenc}
\usepackage{enumitem}
\usepackage[utf8]{inputenc}
\usepackage{multirow}
\usepackage{booktabs, multirow}
\usepackage{color,colortbl}
\usepackage{url}
\usepackage{hyperref}
\usepackage{algorithm}
\usepackage{algorithmic}
\usepackage{caption}
\usepackage{subcaption}
\usepackage{microtype}

\usepackage{inconsolata}

\usepackage{graphicx}

\usepackage{pifont}
\newcommand{\cmark}{\textcolor{green}{\ding{51}}}  
\newcommand{\xmark}{\textcolor{red}{\ding{55}}}   

\newcommand{\dataset}{\textbf{\textit{Defects4Log}}\xspace}

\newcommand{\phead}[1]{\vspace{1mm} \noindent {\bf #1}}
\newcommand{\ulphead}[1]{\vspace{1mm} \noindent \underline {#1}}

\newcommand{\greybox}[1]{
\vspace{0.05cm}
    \begin{tcolorbox}[
        left=2pt, right=2pt, top=2pt, bottom=2pt,
        boxrule=0.2mm,
        leftrule=1mm,
        arc=0mm,
        colframe=black!80!white, 
        colback=black!2!white, 
        colbacktitle=black!50!white 
    ]
    {#1}
    \end{tcolorbox}
\vspace{0.05cm}
}

\definecolor{mygreen}{rgb}{0,0.6,0}
\definecolor{mygray}{rgb}{0.5,0.5,0.5}
\definecolor{mymauve}{rgb}{0.58,0,0.82}

\DeclareCaptionFormat{mylst}{\hrule#1#2#3}
\begin{document}

\title{Defects4Log: Benchmarking LLMs for Logging Code Defect Detection and Reasoning}

\author{
    \IEEEauthorblockN{Xin Wang}
    \IEEEauthorblockA{\textit{The Hong Kong University of Science} \\ \textit{and Technology (Guangzhou)} \\
    Guangzhou, China \\
    xwang496@connect.hkust-gz.edu.cn}
\and
    \IEEEauthorblockN{Zhenhao Li\textsuperscript{*}}
    \IEEEauthorblockA{\textit{York University} \\
    Toronto, Canada \\
    lzhenhao@yorku.ca}
\and
    \IEEEauthorblockN{Zishuo Ding\textsuperscript{*}}
    \IEEEauthorblockA{\textit{The Hong Kong University of Science} \\ \textit{and Technology (Guangzhou)} \\
    Guangzhou, China \\
    zishuoding@hkust-gz.edu.cn}
    
    \thanks{*Corresponding authors.}
}

\maketitle

\begin{abstract}
    Logging code is written by developers to capture system runtime behavior and plays a vital role in debugging, performance analysis, and system monitoring. However, defects in logging code can undermine the usefulness of logs and lead to misinterpretations. Although prior work has identified several logging defect patterns and provided valuable insights into logging practices, these studies often focus on a narrow range of defect patterns derived from limited sources (e.g., commit histories) and lack a systematic and comprehensive analysis. Moreover, large language models (LLMs) have demonstrated promising generalization and reasoning capabilities across a variety of code-related tasks, yet their potential for detecting logging code defects remains largely unexplored.

In this paper, we derive a comprehensive taxonomy of logging code defects, which encompasses seven logging code defect patterns with 14 detailed scenarios. We further construct a benchmark dataset, \dataset, consisting of 164 developer-verified real-world logging defects. 
Then we propose an automated framework that leverages various prompting strategies and contextual information to evaluate LLMs' capability in detecting and reasoning logging code defects. Experimental results reveal that LLMs generally struggle to accurately detect and reason logging code defects based on the source code only. However, incorporating proper knowledge (e.g., detailed scenarios of defect patterns) can lead to 10.9\% improvement in detection accuracy. Overall, our findings provide actionable guidance for practitioners to avoid common defect patterns and establish a foundation for improving LLM-based reasoning in logging code defect detection.

\end{abstract}


\section{Introduction}
\label{sec:intro}

Logging is a fundamental mechanism in software systems, generating runtime execution records (i.e., logs) that provide developers with valuable insights into system behavior and state~\cite{li2024exploring, zhong2025beyond}. These logs are essential for a wide range of software engineering tasks, including debugging, performance analysis, and system monitoring~\cite{yuan_sherlog_2010,barik_bones_2016,milani_comparative_2018,chen_an_empirical_study_leveraging_logs_2019,kim_automatic_2020,li_a_qualitative_study,he_survey_2021,li2023did}.
A logging code typically consists of three main components: a logging level that indicates the severity and importance of the event, a static text message (i.e., logging text) that describes the event, and dynamic variables that provide additional context about the event.
The information captured in the generated logs helps developers better understand and analyze the behavior of their code during execution.

\begin{figure}[t!]
    \centering
    \includegraphics[width=0.5\textwidth]{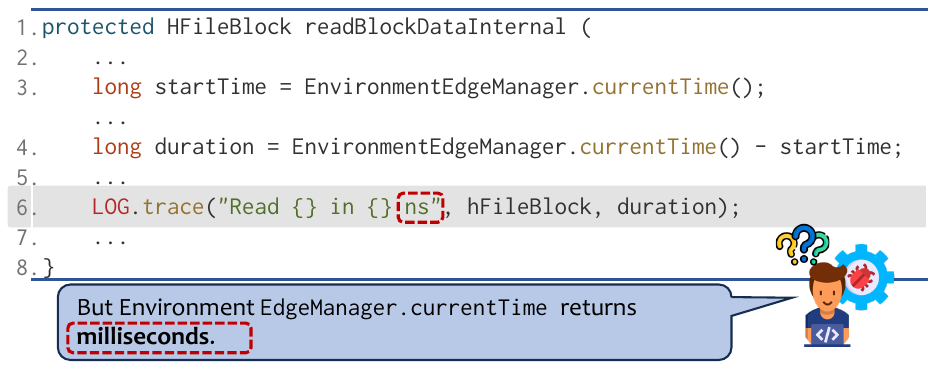} 
    \caption{An issue report from HBase (\href{https://issues.apache.org/jira/browse/HBASE-27099}{HBASE-27099}) highlighting an incorrect logging text on line 6.}
    \label{listing:1}
\end{figure}

However, prior studies~\cite{chen_characterizing_2017,hassani_studying_2018,li_dlfinder_2019,li_a_qualitative_study} have shown that poorly written logging code (i.e., logging code with defects) can significantly reduce the usefulness of logs, or even mislead developers.
As shown in Fig.~\ref{listing:1}, an HBase issue report~\cite{HBASE-27099} describes a defect in the logging code: while the actual return value is measured in milliseconds, the logging text written by developers incorrectly indicates the unit as nanoseconds. This example illustrates how inconsistencies between the semantic meaning of static logging text and its surrounding code context can mislead developers, potentially leading to incorrect assessments of system performance.

Given the severe consequences of logging code defects, several studies have explored their characteristics and developed tools for detecting such defects~\cite{yuan_characterizing_2012,chen_characterizing_2017,hassani_studying_2018,li_dlfinder_2019,ding_temporal_2023,li_are_2023}. While these works provide valuable insights into logging practices, they often focus on a narrow range of defect patterns derived from limited sources (e.g., commit) and lack a systematic and comprehensive analysis of logging code defects. Moreover, most existing detection approaches are rule-based and rely primarily on syntactic and structural analysis, which often overlook the critical factors such as data dependencies that influence runtime behavior. As a result, these techniques are limited to detecting relatively simple and predefined defect patterns.  In contrast, large language models (LLMs) have shown strong generalization and reasoning capabilities across various code-related tasks~\cite{peng2025perfcodegen,chen2025reasoning,peng2025coffe,chen2024code,hu2025assessing,nouriinanloo2024re}, yet their potential for retroactively detecting logging code defects remains unexplored.

To address these limitations, we present the first comprehensive investigation of logging code defects based on three data sources: (1) a systematic literature review, (2) issue reports from tracking systems, and (3) logging-related commits mined from open-source projects.
Through this process, we derive a comprehensive taxonomy comprising seven logging code defect patterns, including previously undocumented ones, such as performance issues caused by logging in hot code paths and sensitive information leakage through logging credential data, along with 14 detailed scenarios where these defects commonly occur. This taxonomy provides a structured foundation for understanding the diverse nature of logging defects. Based on this categorization, we construct the \dataset~benchmark, containing 164 real-world logging code defects. We then systematically evaluate the capabilities of multiple open-source and closed-source LLMs in detecting these defects, exploring the influence of contextual information, domain-specific knowledge, and prompting strategies. Finally, we provide an in-depth analysis of how LLMs' reasoning aligns with developer expectations, highlighting both strengths and areas for improvement.

Through our benchmarking results, we find that detecting defect patterns in logging code remains challenging for LLMs when only the source code is provided as input without additional guidance. In this setting, the average accuracy across various models ranges from 12.4\% to 41.7\%.
This challenge is particularly evident when identifying semantic inconsistencies between logging code and its corresponding code context, where a majority of the average accuracy drops to 0.0\%. This difficulty is due to the complex nature of the task, which goes beyond basic code understanding. LLMs must capture program behavior, track runtime variables, interpret the intent behind the logging text, and reason about their alignment with the surrounding code. Effective detection also requires awareness of project-level logging practices to ensure consistency. These aspects require a deeper, holistic understanding of the source code and its runtime behavior, which current LLMs often lack.

Nonetheless, we find that incorporating domain knowledge (i.e., detailed scenario descriptions of defect patterns) considerably improves LLMs' performance, yielding an absolute improvement of up to 10.9\%. On the other hand, while structural signals from inter-procedural flow (e.g., data and control flow) provide valuable context, current LLMs still struggle to leverage this information effectively. In some cases, the inclusion of these information leads to limited or even degraded performance, underscoring the models' difficulty in capturing nuanced relationships between logging statements and overall program logic. Furthermore, our in-depth analysis of LLM-generated reasoning reveals frequent misalignments between the models' explanations and their final predictions, indicating limitations in reasoning consistency and reliability.

We summarize the contributions of this paper as follows: 

\begin{itemize}

\item We derive a comprehensive taxonomy of logging code defects from multiple sources, including seven defect patterns and 14 detailed scenarios. These findings provide practical insights for practitioners aiming to enhance logging practices by avoiding common defect patterns.

\item We construct \dataset, a real-world benchmark dataset for evaluating the effectiveness of LLMs in logging code defect detection. The replication package of this paper is available~\cite{replication_package}. 

\item We evaluate the effectiveness of several LLMs on the \dataset~benchmark. Experimental results show the effectiveness of our taxonomy and reveal a notable gap between LLMs and human developers. Our findings highlight key limitations and point to future directions for improving LLM-based reasoning in logging code defect detection.
\end{itemize}

\section{Background and Related Work}
\label{sec:related}

\subsection{Improving Logging Code}
Prior work has explored logging defects at varying levels of scope. Some studies have investigated general logging-related issues by analyzing issue reports from two projects~\cite{hassani_studying_2018}, while others have focused on specific aspects such as ``how-to-log'' practices~\cite{chen_characterizing_2017}, temporal inconsistencies between logging and surrounding code~\cite{ding_temporal_2023}, duplicate logging code~\cite{li_dlfinder_2019}, and readability issues~\cite{li_are_2023}. These efforts typically rely on mining commit histories~\cite{chen_characterizing_2017,ding_temporal_2023,li_dlfinder_2019} or conducting developer surveys~\cite{li_are_2023}. More recently, Zhong et al.~\cite{zhong_log_updater} proposed a tool, LogUpdater, which identifies logging defects based on types mined from commit histories, primarily targeting inconsistency and readability issues. 

Our work complements prior studies by introducing a comprehensive taxonomy of logging defects with 14 representative scenarios, derived from multiple sources. We also benchmark LLMs on these patterns to assess their detection and reasoning capabilities.

\subsection{Logging Code Generation}
Prior research on logging code generation has focused on three key questions: where to log, what level to use, and what message to write.
To address where to log, researchers have developed models, often based on deep learning, to analyze code features and recommend optimal logging locations~\cite{zhu_learning_2015, li_where_2020}. To determine the appropriate logging level, deep learning and LLMs have also been used to analyze code context and suggest the correct severity~\cite{li_deeplv_2021,heng2025benchmarking}. Most recently, for generating the logging statement itself, approaches have employed neural machine translation and LLMs to automatically generate log messages and code based on source code context~\cite{ding_logentext_2022, ding_logentext-plus_2023, li_go_2024, zhong2025beyond}. 

All these works focus on the automatic generation of logging code, and our derived taxonomy of logging defects offers guidance for avoiding common anti-patterns. Meanwhile, few have investigated whether the generated logging code exhibits defects, where our taxonomy can serve as a useful reference.

\subsection{Log Analysis}

Log analysis transforms raw logs into structured information to support tasks such as log parsing~\cite{ma_llmparser,xiao2024free}, anomaly detection~\cite{huo2023autolog,xie2024logsd}, fault localization~\cite{chen2021pathidea}, and performance diagnosis~\cite{yao2018log4perf}. Regarding log parsing, researchers have developed various techniques, including parallel systems for scalability, online methods for streaming data, and self-supervised learning models for higher accuracy~\cite{he_characterizing_2018, he_drain_2017, nedelkoski_self-supervised_2021}. Despite their varied goals, all of these approaches rely on the availability of high-quality logs. For anomaly detection, researchers have developed automated systems that transform unstructured log messages into structured patterns to identify execution anomalies, often employing machine learning~\cite{fu_execution_2009, debnath_loglens_2018}. In the field of fault localization, studies have focused on analyzing log classifications and sequences to create contextualized information that guides programmers to the source of an error~\cite{zou_uilog_2016, dobrowolski_log-based_2024}. Within the domain of performance diagnosis, prior work leverages event logs and machine learning to model system behavior and detect performance issues, particularly in domains like HPC~\cite{milani_comparative_2018, kim_automatic_2020}.

Our work contributes at this foundational layer by focusing on detecting defects in logging code. By improving logging quality at the source, we help strengthen the effectiveness of downstream log analysis tasks.

\begin{figure*}[tbp]
    \centering
    \includegraphics[width=1.00\textwidth]{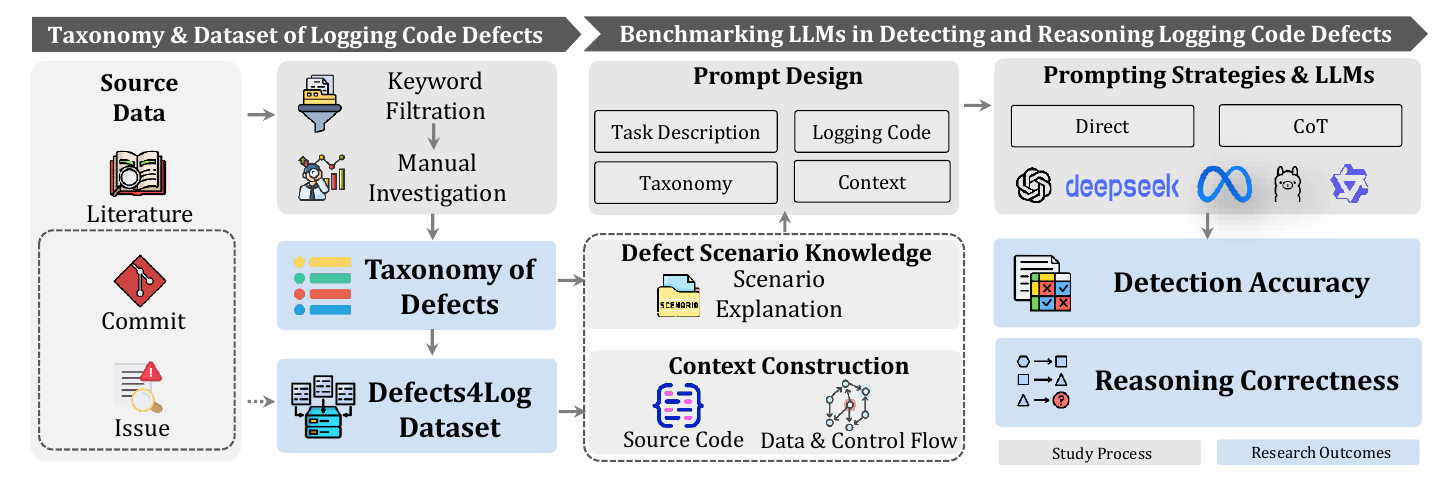} 
    \caption{An overview of our study.}
    \label{fig:overview}
\end{figure*}
\section{Methodology}
\label{sec:methodology}
In this section, we present our methodology for developing a taxonomy of logging code defects, constructing a benchmark dataset of real-world defect instances, and outlining our framework for evaluating the effectiveness of LLMs in detecting these defects. The overall workflow is illustrated in Fig.~\ref{fig:overview}.

\subsection{Deriving the Taxonomy of Logging Code Defects}

To uncover defect patterns in logging code, we conduct a comprehensive analysis of existing literature, issue tracking systems, and commit histories. These sources are strategically selected because they offer complementary perspectives: (1) literature provides established defect classifications, (2) issue reports highlight real-world problems encountered by users and developers, and (3) commit histories reveal defects actively addressed in code along with their fixes. Our objective is to systematically identify and categorize common patterns of logging code defects and to investigate their detailed scenarios. 
\label{label:manual_examination}

\subsubsection{Academic Literature}
We perform a structured literature review using a three-phase methodology. (1) We begin with a search for relevant papers published from 2010 to 2025, on logging code defects using keywords (e.g., ``logging issues'', ``logging anti-patterns'') from major digital libraries (i.e., IEEE Xplore, ACM Digital Library, SpringerLink). (2) We then manually screen the retrieved papers to identify those explicitly addressing logging code defects. (3) To broaden the search scope, we apply a snowballing strategy by reviewing both the references cited by the selected papers and the papers that cite them. This process yields 23 relevant papers. 

\subsubsection{Issue Tracking Systems}
We then collect logging-related defects from issue tracking systems (e.g., JIRA and GitHub). In this process, we base our study on five open-source Java projects: Hadoop, HBase, Hive, Yarn, and ActiveMQ. We select these subject projects because they are widely used in diverse domains (e.g., message brokering, distributed computing, resource management, and databases), are actively maintained, and have been studied in prior research~\cite{hassani_studying_2018,li_dlfinder_2019,ding_temporal_2023,li_are_2023}. We access the issue reports through their respective APIs, which provide structured data including issue titles, descriptions, comments, and associated code changes.

We search for issue reports using keywords (e.g., ``log'', ``logging'')  within issue titles and descriptions, and collect 2,526 logging-related reports. Then, we sample these issues with a 95\% confidence level and 5\% confidence interval, resulting in 334 candidate issue reports. All sampled reports are manually examined by the authors to ensure they describe actual logging code defects. During this process, reports unrelated to logging code defects are filtered out, and 295 issue reports remain for further analysis.

\subsubsection{Commit History}
We also mine software repositories from GitHub to identify commits involving logging code modifications. The process begins with collecting complete commit histories from selected repositories (i.e., Hadoop, HBase, Hive, Camel, and ActiveMQ). To identify relevant commits, we retain 328 commit summaries and descriptions after filtering with relevant keywords and excluding unrelated terms (e.g., ``login'', ``dialog''). Each filtered commit is then manually reviewed to confirm its relevance to logging code defects, and the associated code modifications are extracted for further analysis.

\subsubsection{Manual Examination of Logging Code Defects}
After gathering data from the three sources above, we conduct a detailed manual analysis of the extracted logging-related artifacts to identify and categorize defect patterns. This analysis also involves summarizing the potential scenarios in which each pattern may occur. 
Similar to prior studies~\cite{chen2025understanding,chen2024nlperturbator}, the first author initially analyzes the descriptions and code examples from academic papers, issue reports, commit messages, and associated code changes to derive a set of preliminary defect patterns. These patterns are then reviewed and refined by the second author. In cases of disagreement, a third author participates in the discussion until consensus is reached.

\subsection{Defects4Log Dataset Construction}

Based on our prior manual examination of defect patterns, we construct \dataset, a benchmark for evaluating LLMs in detecting logging code defects. 
Table~\ref{tab:dataset-class-num} presents the distribution of defects in \dataset. 
In total, the dataset comprises 164 real-world logging defect instances across seven detect patterns, collected from commit histories and issue reports.
We discuss each pattern in detail in Section~\ref{sec:results}.

Each instance is labeled with a specific defect pattern based on commit message or issue description, and is accompanied by detailed scenario explanations and relevant contextual information extracted from the source code.
Although the instances are curated from five software systems, the identified defect patterns and our data collection methodology are broadly applicable.

\begin{table}
\centering

\caption{Statistics of \dataset. }
\resizebox{\linewidth}{!} {%
\tabcolsep=12pt
\renewcommand\arraystretch{1.0} 
\begin{tabular}{llc}
\hline
\rowcolor[HTML]{EFEFEF}
\textbf{Abbreviation} & \textbf{Defect Pattern}  & \textbf{Number} \\ \hline
\textbf{RD} & Readability Issues                  & 39     \\
\textbf{VR} & Variable Issues                     & 22     \\
\textbf{LV} & Logging Level Issues                & 26     \\
\textbf{SM} & Semantics Inconsistent with Context & 19     \\
\textbf{SS} & Sensitive Information               & 23     \\
\textbf{IS} & Insufficient Information            & 28     \\
\textbf{PF} & Performance Issues                  & 7     \\
\midrule
\multicolumn{2}{r}{\textit{Total}}               & 164  \\
\bottomrule
\end{tabular}
}
\label{tab:dataset-class-num}
\end{table}

\subsection{Framework for Benchmarking LLMs in Detecting Logging Code Defects}

In this section, we introduce our framework for evaluating the extent to which LLMs can assist developers in detecting logging code defects. As shown in Fig.~\ref{fig:overview},  the framework consists of four components: \textit{Context Construction}, \textit{Defect Scenario Knowledge}, \textit{Prompt Design}, and \textit{Prompting Strategy}. We describe each component in detail below.

\subsubsection{Context Construction}

Contextual information has been widely used to enhance the reasoning capabilities of LLMs in various code-related tasks~\cite{yun_project-specific_2024, lomshakov_proconsul_2024, khan_code_2024, su_context-aware_2024, zhang_context-aware_2024, chen2025reasoning}. For logging code, context is especially critical, as logging code is closely tied to the execution behavior of surrounding code and the runtime state of relevant variables. Accurately interpreting logging code therefore requires a clear understanding of its surrounding context and its relationship to overall program behavior.

In our framework, we focus on three types of contextual information: (1) direct context, referring to the source code of the function that contains the logging code; and two forms of inter-procedural flow context: (2) control flow, which captures the execution paths leading to the logging code; and (3) data flow, which traces how data influences the variables referenced within the logging code.

\noindent\textbf{Source code of the function.} The primary type of context for logging code is the source code of the function in which the logging code resides. This context has been widely utilized in prior work on logging-related tasks~\cite{ding_logentext_2022, ding_logentext-plus_2023, li_go_2024, li_studying_duplicate}. It captures the immediate surroundings of the logging code, such as declarations, that directly interact with or affect the logging code. Therefore, we include this function-level code as the basic form of contextual information in our framework.

\noindent\textbf{Control flow.}  We also consider the program's control flow, which captures the potential execution sequences and decision points. This can help isolate the specific conditions or paths under which a logging code is executed by filtering out irrelevant code paths. In our framework, control flow context is defined by two specific paths relative to the logging code: a backward path, capturing the code executed before the logging code, and a forward path, representing the code executed afterward. The scope of these paths is controlled by a hyperparameter that limits the maximum length traced in each direction from the logging code.

\noindent\textbf{Data flow.} We also incorporate data flow context, as it directly impacts the runtime state of the variables captured in logging code. To construct this context, we first identify all variables referenced in the logging code. We then perform a backward data-flow analysis, starting from each usage and tracing dependencies in reverse to their definitions. The analysis proceeds until it reaches a data source, such as constant assignments, method parameters, or return values from function calls. In the case of method invocations, the analysis can either stop at the call site or be extended into the callee function. This process allows us to recover the computation chain that shapes the runtime state of each logged variable.

\subsubsection{Defect Scenario Knowledge} 

Although LLMs show strong performance across diverse tasks, they often struggle with domain-specific issues without explicit contextual guidance~\cite{kojima_large_2022,hu_ella_2024}. To systematically assess this, our benchmark incorporates defect scenario knowledge into the prompting process. Specifically, we provide scenario names and their corresponding explanations, which define the characteristics of each defect pattern and serve as structured guidance for evaluating the LLMs' ability to leverage such knowledge.

\subsubsection{Prompt Design}
We design several prompt templates to evaluate LLMs under different input configurations. Each template is built around a core structure that includes the task description (i.e., decide from \textit{``no defect''}, or any one of the defect patterns), the source code context (i.e., the function containing the logging code), the target logging code, and the taxonomy of defect patterns. Variants of this base template additionally include inter-procedural flow context (i.e., control flow and data flow), domain knowledge (i.e., scenario descriptions), or both. These prompts are intended to guide the LLM in predicting the correct defect pattern and generating an explanation of its reasoning process.

\subsubsection{Prompting Strategy}

We experiment with two prompting strategies: \textbf{Direct} prompting, and \textbf{Chain of Thought (CoT)} prompting~\cite{wei_chain--thought_2022}. Both strategies are applied using the prompt templates described in the previous section.

\section{Experiment Setup}
\label{section:experiment_setup}
\subsection{Research Questions}
\noindent We study the following research questions:

\begin{itemize}
    \item \textbf{RQ1:} How prevalent are logging code defects in open-source projects?
    \item \textbf{RQ2:} How is the effectiveness of different LLMs in detecting logging code defects?
    \item \textbf{RQ3:} How can LLMs correctly reason the detected logging code defects?

\end{itemize}

\subsection{Evaluation Metrics}

The detection task is formulated as a multi-class classification problem, where LLMs are expected to determine whether the given input corresponds to one of the defect patterns or is non-defective. A detection result is considered correct if the identified pattern matches the ground truth. We compute the accuracy for each defect pattern individually and report the macro-average accuracy. This enables a more fine-grained evaluation of the model's performance across different patterns of logging defects.

$$
\text{Accuracy}_{\text{pattern}} = \frac{\text{Number of Correctly Detected Instances}}{\text{Total Instances of the Pattern}}
$$

\subsection{Studied LLMs and Implementation Details.}
In our experiments, we study four widely used LLMs, including both open-source and closed-source ones. For open-source models, we select \texttt{Qwen2.5-72B}~\cite{qwen_qwen25_2025}, \texttt{DeepSeek-R1}~\cite{deepseek-ai_deepseek-r1_2025}, and \texttt{Llama3.3}~\cite{grattafiori_llama_2024}. For the closed-source model, we choose \texttt{GPT-4o}~\cite{opeai_gpt4}.

All experiments are conducted on a Linux server running Ubuntu 20.04.6 LTS, equipped with an AMD 32-core processor, 1TB RAM, and eight NVIDIA A6000 GPUs.
To extract inter-procedural flow information from the source code, we use SciTools Understand~\cite{understand}.

\section{Results}
\label{sec:results}

In this section, we discuss the results of each RQ.

\subsection{\textbf{RQ1:} Prevalence of Logging Code Defects}

\begin{table}[t]
\centering
\caption{Taxonomy of Logging Code Defects in \dataset (RQ1)  }

\resizebox{\linewidth}{!} {%
\tabcolsep=1pt
\begin{tabular}{l|l}
\hline
\rowcolor[HTML]{EFEFEF}
\textbf{Defect Patterns} & \textbf{Scenario Name}   \\
\hline

 \multirow{3}{*}{\textbf{RD:} Readability Issues} 
   & \textbf{RD-1:} Complicated domain-specific terminology   \\
 & \textbf{RD-2:} Non-standard language used    \\
 & \textbf{RD-3:} Poorly formatted or unclear messages    \\

\midrule

 \multirow{2}{*}{\textbf{VR:} Variable Issues
}
   & \textbf{VR-1:} Incorrect variable value logging    \\
  & \textbf{VR-2:} Placeholder–value mismatch    \\

\midrule

  \textbf{LV:} Logging Level Issues & \textbf{LV-1: } Improper verbosity level      \\
\midrule

 \multirow{3}{*}{\textbf{SM:} Semantics Inconsistent}
   & \textbf{SM-1:} Wrong unit or metric label     \\
  & \textbf{SM-2:} Message text does not match the code   \\
 & \textbf{SM-3:} Misused variables in the message    \\

\midrule

 \multirow{2}{*}{\textbf{SS:} Sensitive Information}
   & \textbf{SS-1:} Credentials logged in plain text  \\
  & \textbf{SS-2:} Dumping whole objects without scrubbing     \\

\midrule

  \textbf{IS:} Insufficient Information & \textbf{IS-1:} Insufficient information   \\
\midrule

  \multirow{2}{*}{\textbf{PF:} Performance  Issues} & \textbf{PF-1:} Logging on hot path.  \\ 
 & \textbf{PF-2:} Costly string operations    \\

\bottomrule
\end{tabular}
}
\label{tab:pattern}
\end{table}

Through our comprehensive investigation (cf. Section~\ref{label:manual_examination}), we develop a taxonomy consisting of seven logging code defect patterns and 14 detailed scenarios in which these defects commonly occur. Table~\ref{tab:pattern} presents an overview of these patterns. In the following, we elaborate on each defect pattern, describe its corresponding scenarios with the number of instances in \dataset inside \textit{``()''}, and provide real-world examples.

\phead{Readability Issues (RD).} This defect pattern includes issues in logging code that reduce the clarity of the produced log messages, making them harder for practitioners to interpret and for automated tools to parse.

\ulphead{Defect Scenarios:}

\begin{itemize}[leftmargin=*, label=•]
    
    \item \textbf{RD-1} \textit{Complicated domain-specific terminology (2):} Complex domain terms that may reduce message clarity for broader audiences. 
    \item \textbf{RD-2} \textit{Non-standard language used (15):} Grammatical errors or informal phrasing that affect message professionalism and clarity.
    \item \textbf{RD-3} \textit{Poorly formatted or unclear messages (22):} Poorly structured messages that impede comprehension and hinder debugging. 

\end{itemize}

\ulphead{Real-world Example:} HBASE-24367~\cite{HBASE-24367} presents a case where the average elapsed time is logged in nanoseconds, which one comment describes as non-intuitive or even ``annoying''. The issue reporter also notes that such high precision is unnecessary for averages. In the fix for similar issues, developers convert the time unit to milliseconds, with the commit message stating ``log elapsed timespan in a human-friendly format''. Detecting such issues requires LLMs to reason about return value ranges via call chains and interpret the semantics of logging text (e.g., recognizing that ``average'' implies high precision is unnecessary in such situations).
\begin{lstlisting}[
    language=Java,
    label=list:RD,
    captionpos=t,
    breaklines=true,
    basicstyle=\ttfamily\footnotesize,
    tabsize=2,               
    breakatwhitespace=true,  
    breakautoindent=true,   
    breakindent=10pt,                                                                 
    keepspaces=true,      
    columns=flexible,      
    showstringspaces=false 
]
// HBASE-24367
LOG.info("{} average execution time: {} ns.", getName(), (long)(timeMeasurement.getAverageTime()));
\end{lstlisting}

\phead{Variable Issues (VR).} This defect pattern refers to logging incorrect or mismatched variables that may cause confusion or mislead developers during debugging.

\ulphead{Defect Scenarios:} 

\begin{itemize}[leftmargin=*, label=•]
    \item \textbf{VR-1} \textit{Incorrect variable value logging (3):} Logging a null or raw address instead of the actual value that may cause crashes or mislead developers. 
    \item \textbf{VR-2} \textit{Placeholder–value mismatch (19):} Mismatched placeholders and variables that lead to confusing or misleading outputs.
\end{itemize}

\ulphead{Real-world Example:} A commit in AutoMQ~\cite{AutoMQ-Commit-e6f8ca8} illustrates a clear case of a placeholder-argument mismatch. As shown in the listing below, the logging code contains three placeholders, but only two arguments are provided, resulting in the omission of the \texttt{epoch} value. The issue has been resolved by adding the missing \texttt{epoch} variable. Detecting such patterns requires models to accurately align placeholders with their corresponding arguments, while fixing them demands deep contextual reasoning to infer the developer’s intent and identify the appropriate variable from the surrounding code.
\begin{lstlisting}[
    language=Java,
    label=list:VR,
    captionpos=t,
    breaklines=true,
    basicstyle=\ttfamily\footnotesize,
    tabsize=2,              
    breakatwhitespace=true,  
    breakautoindent=true,    
    breakindent=10pt,                                                               
    keepspaces=true,       
    columns=flexible,      
    showstringspaces=false 
]
// AutoMQ-Commit-e6f8ca8
log.trace("tryAppend(nodeId={}, epoch={}): the given node id does not match the current leader id of {}.", nodeId, leader.nodeId());
\end{lstlisting}

\phead{Logging Level Issues (LV).} This pattern captures issues where the assigned logging level is inconsistent with the severity or importance of the event.

\ulphead{Defect Scenarios:} 
\begin{itemize}[leftmargin=*, label=•]
    \item \textbf{LV-1} \textit{Improper verbosity level (26):} The verbosity level that does not accurately reflect the seriousness of the message. 
\end{itemize}

\ulphead{Real-world Example:} Issue HDFS-15045~\cite{HDFS-15045} demonstrates an inappropriate logging level where an exception is logged using \texttt{INFO}, as shown in the code below.
Using the \texttt{INFO} level for an exception is risky because the message can be easily overlooked. A developer on the issue report confirmed this risk, noting that clients often set their levels to \texttt{WARN}. In this common scenario, the logs for this exception would be completely hidden, preventing users from understanding the reason for a critical pipeline failure.
Detecting this issue requires the model to go beyond syntax and understand implicit software engineering conventions (e.g., avoiding logging exceptions at the \texttt{INFO} level), as well as reason about the severity of the system behavior captured by the corresponding logging code.

\begin{lstlisting}[
    language=Java,
    label=list:LV,
    captionpos=t,
    breaklines=true,
    basicstyle=\ttfamily\footnotesize,
    tabsize=2,               
    breakatwhitespace=true,  
    breakautoindent=true,    
    breakindent=10pt,                                                               
    keepspaces=true,      
    columns=flexible,     
    showstringspaces=false 
]
// HDFS-15045
LOG.info("Exception in createBlockOutputStream " + this, ie);
\end{lstlisting}

\phead{Semantics Inconsistent (SM).} This pattern encompasses logging code that is semantically inconsistent with the actual behavior or state of the surrounding code.

\ulphead{Defect Scenarios:} 

\begin{itemize}[leftmargin=*, label=•]
    \item \textbf{SM-1} \textit{Wrong unit or metric label (1):} Incorrect units or labels that may mislead interpretation. 
    \item \textbf{SM-2} \textit{Message text doesn’t match the code’s action or state (12):} The log message that misrepresents the actual code behavior or logic. 
    \item \textbf{SM-3} \textit{Misused variables in the message (6):} Incorrect variables that may mislead users.
\end{itemize}

\ulphead{Real-world Example:} 
Issue YARN-6951~\cite{YARN-6951} demonstrates a semantics inconsistent defect where the logging text contradicts the actual program logic. As shown in the listing below, the code logs ``Resource handler chain enabled = true'' when the \texttt{resourceHandlerChain} is \texttt{null}, which is misleading because \texttt{null} indicates an inactive or non-existent state. A developer has confirmed in the issue report that this was incorrect and clarified that the logs should display ``true'' only when the chain is not \texttt{null}. To detect it, a model must understand the natural language semantics of the word ``enabled'' and recognize its logical contradiction with the programming convention where null implies a disabled state, a reasoning process that requires aligning natural language with code semantics.

\begin{lstlisting}[
    language=Java,
    label=list:SM,
    captionpos=t,
    breaklines=true,
    basicstyle=\ttfamily\footnotesize,
    tabsize=2,              
    breakatwhitespace=true,  
    breakautoindent=true,   
    breakindent=10pt,                                                               
    keepspaces=true,      
    columns=flexible,     
    showstringspaces=false 
]
// YARN-6951
LOG.debug(\"Resource handler chain enabled = \" + (resourceHandlerChain == null));
\end{lstlisting}

Similarly, Fig.~\ref{listing:1} shows a case of inconsistency~\cite{HBASE-27099} where the variable \texttt{duration} returns a value in milliseconds, while the logging text incorrectly labels it as ``ns'' (nanoseconds). Although this problem appears similar to the earlier example~\cite{HBASE-24367} categorized under Readability Issues, which also involves a unit correction, it reflects a completely different defect pattern. While the previous case affects log clarity for human readers, this example concerns the semantic correctness of the logged information. Detecting it requires the model to trace the data flow and reason about the true unit of the variable, highlighting the need for a deeper understanding of program behavior.

\phead{Sensitive Information (SS).} This defect pattern captures scenarios where generated logs expose private or confidential data, a critical issue that has been overlooked in prior research on logging defects.

\ulphead{Defect Scenarios:} 

\begin{itemize}[leftmargin=*, label=•]
    \item \textbf{SS-1} \textit{Credentials logged in plain text (7):} Logging code that exposes private or confidential data (e.g., password, token) in plain text during system execution, potentially leading to security and privacy risks. 
    \item \textbf{SS-2} \textit{Dumping whole objects / configs without scrubbing (16):} Sensitive information may be embedded within objects or configuration data, which can be inadvertently exposed if not properly scrubbed.
\end{itemize}

\ulphead{Real-world Example:} 
Issue HIVE-20796~\cite{HIVE-20796} demonstrates the risk of logging potentially sensitive information. As shown in the listing below, the variable \texttt{confVal} is logged directly, which poses a security risk because it can contain a JDBC URL. A developer on the issue report has elaborated on this risk, explaining that such URLs can embed sensitive credentials like passwords and therefore must be ``masked out'' before being logged. Detecting this defect requires the model not only to recognize what types of information are considered sensitive, but also to infer the potential contents of generic variables like \texttt{confVal}, which may hold confidential data. Additionally, it must understand the security principle of data masking, a process that extends beyond code syntax and draws on real-world security practices to properly address such issues.

\begin{lstlisting}[
    language=Java,
    label=list:SS,
    captionpos=t,
    breaklines=true,
    basicstyle=\ttfamily\footnotesize,
    tabsize=2,              
    breakatwhitespace=true,  
    breakautoindent=true,   
    breakindent=10pt,                                                                
    keepspaces=true,      
    columns=flexible,     
    showstringspaces=false 
]
// HIVE-20796
LOG.debug("Overriding {} value {} from jpox.properties with {}", varName, prevVal, confVal);
\end{lstlisting}

\phead{Insufficient Information (IS).} This pattern encompasses logging code that lacks essential details needed to understand or diagnose events.

\ulphead{Defect Scenarios:}

\begin{itemize}[leftmargin=*, label=•]
    \item \textbf{IS-1} \textit{Insufficient information (28):} This issue typically arises when an error occurs and developers need more contextual information for diagnosis, but the logs lack key error details or essential event context.
\end{itemize}

\ulphead{Real-world Example:} Issue HDFS-17310~\cite{HDFS-17310} illustrates the logging code with insufficient context for effective debugging. As shown in the listing below, the original error logging code states that a plan submission failed but provides no specific identifiers. To make troubleshooting more convenient, as noted by the developer, the logging code has been enhanced by adding crucial details: the \texttt{planFile} and \texttt{planId}. Detecting such defects is challenging, as it requires the model to assess the diagnostic quality of logs and reason about whether they lack sufficient contextual information for effective troubleshooting.

\begin{lstlisting}[
    language=Java,
    label=list:IS,
    captionpos=t,
    breaklines=true,
    basicstyle=\ttfamily\footnotesize,
    tabsize=2,              
    breakatwhitespace=true, 
    breakautoindent=true,   
    breakindent=10pt,                                                               
    keepspaces=true,       
    columns=flexible,      
    showstringspaces=false 
]
// HDFS-17310
LOG.error("Disk Balancer - Executing another plan, submitPlan failed.");
\end{lstlisting}

\phead{Performance Issues (PF).} This defect pattern refers to logging code that may negatively affect system performance due to where or how they are executed.

\ulphead{Defect Scenarios:} 

\begin{itemize}[leftmargin=*, label=•]
    \item \textbf{PF-1} \textit{Logging on hot path (6):} Logging in tight loops or latency-sensitive code that impacts system performance. 
    \item \textbf{PF-2} \textit{Costly string operations (1):} Costly string concatenation or formatting that increases CPU or memory usage.   
\end{itemize}

\ulphead{Real-world Example:} Issue HIVE-25794~\cite{HIVE-25794} reports a performance issue caused by using string concatenation inside a frequently executed \texttt{LOG.debug()} call. In the issue report, a developer mentions that this leads to memory pressure when processing a large number of files, as the string construction occurs even when the logging level is set to \texttt{INFO}. The issue has been fixed by replacing the concatenation with a parameterized logging approach using placeholders. Detecting such issues requires reasoning about the cost of string operations and their impact in performance-critical code.

\begin{lstlisting}[
    language=Java,
    label=list:PF,
    captionpos=t,
    breaklines=true,
    basicstyle=\ttfamily\footnotesize,
    tabsize=2,              
    breakatwhitespace=true, 
    breakautoindent=true,   
    breakindent=10pt,                                                                
    keepspaces=true,       
    columns=flexible,      
    showstringspaces=false 
]
// HIVE-25794
LOG.debug("Found spec for " + path + " " + otherPart + " from " + pathToPartInfo);
\end{lstlisting}

\greybox{\textbf{RQ1 Summary:} The taxonomy of \dataset summarizes seven patterns of logging code defects and 14 real-world defect scenarios. We find that the summarized logging code defects are prevalent in real-world projects, with some defects (e.g., Readability Issues) appearing more commonly in our studied projects.}

\subsection{\textbf{RQ2:} Effectiveness of LLMs on Defects4Log}

\begin{table*}
\centering 
\caption{Accuracy (\%) of LLMs on \dataset (RQ2).}
    \resizebox{\linewidth}{!} {
        \tabcolsep=18pt
        \renewcommand\arraystretch{0.72}
\begin{tabular}{l|l|ccccccc|l}
\hline
 
\rowcolor[HTML]{EFEFEF} \textbf{\rule{0pt}{10pt}Model} & \textbf{Setting} & \textbf{RD} & \textbf{VR} & \textbf{LV} & \textbf{SM} &\textbf{SS} & \textbf{IS}   & \textbf{PF}  & \textit{Average} \\

\hline

\multirow{7}{*}{\textit{GPT-4o}} 
&\textit{Direct}                &46.2 &9.1  &3.8  &0.0     &34.8 &14.3  &14.3    &17.5  \\
&\textit{Direct\textit{+K}}     &41.0 &36.4 &15.4 &0.0     &52.2 &39.3  &14.3    &28.4 (+10.9) \\
&\textit{Direct\textit{+I}}     &46.2 &0.0  &3.8  &0.0     &43.5 &14.3  &28.6    &19.5 (+2.0) \\
&\textit{Direct\textit{+K+I}}   &41.0 &27.3 &7.7  &0.0     &52.2 &39.3  &0.0     &23.9 (+6.4)  \\
\cmidrule{2-10}
&\textit{CoT}                   &43.6 &13.6 &3.80 &0.0     &47.8 &14.3  &14.3    &19.6  \\
&\textit{CoT\textit{+K}}        &41.0 &27.3 &15.4 &5.3     &52.2 &\textcolor{red}{\textbf{50.0}}  &0.0     &27.3 (+7.7)  \\
&\textit{CoT\textit{+I}}        &43.6 &4.5  &19.2 &5.3     &47.8 & 7.1  &14.3    &20.2 (+0.6)  \\
&\textit{CoT\textit{+K+I}}      &41.0 &27.3 &7.7  &15.8    &52.2 &39.3  &0.0     &26.2 (+6.6)  \\

\midrule

\multirow{7}{*}{\textit{Llama3.3}} 
&\textit{Direct}              &51.3 &0.0  &3.8  &0.0     &52.2  &3.6   &0.0   &15.8  \\
&\textit{Direct\textit{+K}}   &61.5 &4.5  &0.0  &0.0     &43.5  &17.9   &0.0   &18.2 (+2.4)  \\
&\textit{Direct\textit{+I}}   &53.8 &0.0  &0.0  &0.0     &47.8  & 3.6   &0.0   &15.0 (-0.8)  \\
&\textit{Direct\textit{+K+I}} &\textcolor{red}{\textbf{69.2}} &4.5  &0.0  &0.0     &34.8  & 7.1   &0.0   &16.5 (+0.7)  \\
\cmidrule{2-10}
&\textit{CoT}                 &51.3 &0.0 &3.8   &0.0     &47.8  & 3.6   &0.0    &15.2  \\
&\textit{CoT\textit{+K}}      &48.7 &4.5 &0.0   &0.0     &30.4  &14.3   &0.0    &14.0 (-1.2) \\
&\textit{CoT\textit{+I}}      &46.2 &0.0 &0.0   &0.0     &52.2  & 7.1   &0.0    &15.1 (-0.1)\\
&\textit{CoT\textit{+K+I}}    &\textcolor{red}{\textbf{69.2}} &4.5 &0.0   &0.0     &34.8  & 7.1   &0.0    &16.5 (+1.3)  \\

\midrule

\multirow{8}{*}{\textit{DeepSeek-R1}}
& \textit{Direct}          & 56.4 & 22.7 &\textcolor{red}{\textbf{19.2}} & 31.6 &\textcolor{red}{\textbf{65.2}} & 39.3 & 57.1 & 41.7 \\
 & \cellcolor[HTML]{FDE9E9}\textit{Direct+K} & 56.4 & \textcolor{red}{\textbf{68.2}} & 15.4 & 42.1 &\textcolor{red}{\textbf{65.2}} & 28.6 & 57.1 &\cellcolor[HTML]{FDE9E9}\textcolor{red}{\textbf{47.6}} (+5.9) \\ 
& \textit{Direct+I}        & 48.7 & 22.7 & 15.4 & 47.4 & 56.5 & 42.9 & 57.1 & 41.5 (-0.2) \\
& \textit{Direct+K+I}      & 46.2 & 40.9 &\textcolor{red}{\textbf{19.2}} &\textcolor{red}{\textbf{52.6}} & 56.5 & 32.1 & 71.4 & 45.6 (+3.9) \\
\cmidrule{2-10}
& \textit{CoT}             & 53.8 & 13.6 &\textcolor{red}{\textbf{19.2}} & 21.1 & 56.5 & 32.1 & 57.1  & 36.2 \\
& \textit{CoT+K}           & 53.8 & \textcolor{red}{\textbf{68.2}} &  7.7 &  0.0 & 56.5 & 28.6 &\textcolor{red}{\textbf{100.0}} & 45.0 (+8.8) \\
& \textit{CoT+I}           & 43.6 & 22.7 &  7.7 & 42.1 & 52.2 & 32.1 & 71.4  & 38.8 (+2.6)\\
& \textit{CoT+K+I}         & 46.2 & 40.9 &\textcolor{red}{\textbf{19.2}} &\textcolor{red}{\textbf{52.6}} & 56.5 & 32.1 & 57.1  & 43.5 (+7.3) \\

\midrule
\multirow{7}{*}{\textit{Qwen2.5-72B}} 
&\cellcolor[HTML]{E6F2FF}\textit{Direct}                &23.1 &9.1 &7.7 &0.0     &21.7 &10.7   &14.3      &\cellcolor[HTML]{E6F2FF}12.4  \\
&\textit{Direct\textit{+K}}     &46.2 &9.1 &3.8 &0.0     &34.8 &14.3   &14.3      &17.5 (+5.1)  \\
&\textit{Direct\textit{+I}}     &35.9 &0.0 &3.8 &0.0     &17.4 & 3.6   &28.6      &12.8 (+0.4)  \\
&\textit{Direct\textit{+K+I}}   &46.2 &9.1 &3.8 &0.0     &26.1 &17.9   &14.3      &16.8 (+4.4)  \\
\cmidrule{2-10}
&\textit{CoT}               &30.8 & 9.1  &3.8 &0.0     &21.7 &17.9    &28.6    &16.0  \\
&\textit{CoT\textit{+K}}    &59.0 &13.6  &0.0 &0.0     &21.7 &14.3    & 0.0    &15.5 (-0.5)  \\
&\textit{CoT\textit{+I}}    &38.5 & 4.5  &3.8 &0.0     &26.1 & 3.6    &14.3    &13.0 (-3.0)  \\
&\textit{CoT\textit{+K+I}}  &46.2 & 9.1  &3.8 &0.0     &26.1 &17.9    &14.3    &16.8 (+0.8)  \\

\bottomrule
\end{tabular}
}

\label{tab:effectiveness}
\end{table*}

In this RQ, we aim to benchmark the effectiveness of current LLMs in detecting logging code defects.

\phead{Approach.}
Using the dataset constructed in RQ1 and the accuracy metric defined in Section~\ref{section:experiment_setup}, we evaluate LLM performance across three key dimensions: (1) the choice of LLM, (2) the prompting strategy used, and (3) the influence of different types of input context information.

 Table~\ref{tab:effectiveness} presents the results of this RQ for each pattern.  The \textit{Setting} column distinguishes between \texttt{Direct} (i.e., direct prompting) and \texttt{CoT} (i.e., chain-of-thought prompting) with different input contexts. The modifier \texttt{+K} indicates the inclusion of knowledge of defect scenarios, and \texttt{+I} indicates the inclusion of inter-procedural information.
For each setting, we report individual results per pattern and the averaged accuracy across patterns. For each pattern, the highest accuracy is highlighted in \textcolor{red}{\textbf{bold and red}}. The prompting strategy with the highest average accuracy is highlighted in \colorbox[HTML]{FDE9E9}{red} in its cell, while the lowest is highlighted in \colorbox[HTML]{E6F2FF}{blue}.

\phead{Comparison of LLMs.}
Overall, \texttt{Deepseek-R1} consistently achieves the best performance across the majority of prompting settings. Specifically, its \texttt{Direct+K} setting yields the highest average accuracy (i.e., 47.6\%), with the highest results on patterns such as VR (i.e., 68.2\%), and SS (65.2\%). Its \texttt{Direct+K+I} setting follows with an average accuracy of 45.6\%. \texttt{GPT-4o} performs moderately well, with its best average accuracy of 27.3\% under \texttt{CoT+K}. \texttt{Llama3.3} achieves high accuracy in RD (i.e., 69.2\%) under \texttt{Direct+K+I} and \texttt{CoT+K+I} (69.2\%) but almost completely ineffective in some of the patterns (e.g., 0.0\% in SM and PF). \texttt{Qwen2.5-72B} achieves the lowest average accuracy (i.e., 12.4\%) in the setting of \texttt{Direct}.

\phead{Comparison of Prompting Strategies.}
We find that CoT does not obviously outperform \texttt{Direct} across different models and patterns. For instance, in \texttt{GPT-4o}, the average accuracy of \texttt{CoT+K+I} is 26.2\%, only slightly higher than \texttt{Direct+K+I} (i.e., 23.9\%), while some CoT variants (e.g., \texttt{CoT+K} = 27.3\%) perform worse than their direct settings (e.g., \texttt{Direct+K} = 28.4\%). A similar trend is observed in \texttt{Llama3.3}, where \texttt{Direct+K} (i.e., 18.2\% on average) outperforms all CoT-based settings. Even in the best-performing model, \texttt{Deepseek-R1}, the improvement from CoT prompting is marginal and sometimes reversed. For example, \texttt{Direct+K} achieves a higher average (i.e., 47.6\%) than \texttt{CoT+K} (45.0\%). These results suggest that \texttt{CoT} prompting may not provide noticeable improvements for tasks involving detecting logging code defects, especially when the models already have sufficient capabilities to reason directly from enriched input. Instead, it is possible that \texttt{CoT} may introduce unnecessary verbosity or distract from the precise detection of defects related to logging code.

\phead{Impact of Input Context.}
As shown in Table~\ref{tab:effectiveness}, we also report the relative change in average performance when augmenting the prompt with detailed knowledge of pattern scenarios (\texttt{+K}) and/or inter-procedural information (\texttt{+I}). Overall, adding detailed knowledge of pattern scenarios (\texttt{+K}) tends to consistently improve the accuracy across various prompting settings and LLMs. For example, \texttt{GPT-4o}'s average accuracy increases from 17.5\% for \texttt{Direct} to 28.4\% for \texttt{Direct+K}, and \texttt{Deepseek-R1}'s performance improves from 41.7\% for \texttt{Direct} to 47.6\% for \texttt{Direct+K}. Similarly, \texttt{Qwen2.5-72B} has an improvement from 12.4\% to 17.5\% for \texttt{Direct+K}. This indicates that providing LLMs with concrete scenarios of the logging code defect patterns may help them better understand the meaning of those defects and perform the detection. 

In contrast, incorporating inter-procedural information (\texttt{+I}) yields mixed results. While it enhances performance in some patterns (e.g., improved PF in \texttt{Deepseek-R1} under \texttt{Direct+K+I}), it does not provide consistent gains. In some cases, the inclusion of inter-procedural information leads to performance drops, such as in \texttt{Llama3.3} where \texttt{Direct+I} (i.e., 15.0\%) performs worse than \texttt{Direct} (i.e., 15.8\%). These observations suggest that although \texttt{+I} provides useful contextual cues, its effectiveness depends on how well the underlying LLMs can interpret and leverage static code structures. Without proper alignment, it may introduce additional noise rather than benefit the reasoning process.

\greybox{\textbf{RQ2 Summary: }We find that \texttt{Deepseek-R1} generally outperforms other LLMs (e.g., \texttt{Direct+K} achieves an average accuracy of 47.6\%). In terms of prompting strategies, \texttt{CoT} does not demonstrate clear improvements over \texttt{Direct} prompting. Additionally, including detailed knowledge of pattern scenarios (\texttt{+K}) leads to noticeable performance gains, while incorporating inter-procedural information (\texttt{+I}) shows mixed effects and sometimes decreases performance.}

\subsection{\textbf{RQ3:} Capability of Reasoning the Results}
In RQ2, we focus on the final prediction, specifically, whether the LLMs can correctly detect the defect patterns in logging code. However, relying solely on the final answer may overlook valuable insights embedded in the model's reasoning process. This is critical because, even if an LLM correctly detects a defect, it may fail to provide a coherent or accurate explanation of why the code is defective. Such reasoning gaps can lead to misunderstandings or negatively affect downstream tasks, such as defect fixing, thereby limiting the model's practical utility in assisting developers. Therefore, in this research question, we investigate the LLM's ability to generate accurate and meaningful reasoning behind its predictions.

\phead{Approach.} To answer this RQ, we first select cases generated by the best-performing LLM, \texttt{Deepseek-R1} with \texttt{Direct} prompt augmented by domain knowledge (i.e., \texttt{\texttt{Direct+K}}), and the worst-performing LLM, \texttt{Qwen2.5-72B} with the \texttt{\texttt{\texttt{Direct}}} prompt (i.e., \texttt{\texttt{Direct}}), as they represent two distinct performance extremes.
We then manually evaluate the reasoning provided by each model by comparing it against the ground truth, which is derived by the authors based on code changes, issue descriptions, developer discussions, and commit messages, as well as the surrounding source code. Two authors independently review and label the reasoning correctness, achieving substantial agreement with a Cohen’s Kappa score of 0.88~\cite{cohen_coefficient_1960}.

\begin{table*}
\centering

\caption{Evaluation of LLM Reasoning: Correctness of Defect Pattern Detection and Reasoning (RQ3)
}
    \resizebox{\linewidth}{!} {
        \tabcolsep=12pt 
        \renewcommand\arraystretch{0.95} 
\begin{tabular}{l|cc|ccccccc|c} 
\hline

\rowcolor[HTML]{EFEFEF} \textbf{\rule{0pt}{10pt}Model \& Setting} & \textbf{Pattern}  & \textbf{Reason} & \textbf{RD} & \textbf{VR} & \textbf{LV} & \textbf{SM} &\textbf{SS} & \textbf{IS}    & \textbf{PF} & \textit{Average}  \\

\hline

\multirow{4}{*}{\textit{DeepSeek-R1 \& Direct+K}} 
&\cmark  &\cmark       &55.3 &69.6 &11.5  &42.1  &65.2 &28.6  &57.1 &45.7     \\
&\cmark   &\xmark     &2.6 &0.0 &3.8  &0.0  &0.0 &0.0  &0.0  &1.2   \\
&\xmark   &\cmark     &0.0 &0.0 &0.0  &0.0  &0.0 &0.0  &0.0  &0.0  \\
&\xmark   &\xmark    &42.1 &30.4 &84.6  &57.9  &34.8 &71.4  &42.9  &53.0   \\
\midrule
\multirow{4}{*}{\textit{Qwen2.5-72B \& Direct}}
&\cmark   &\cmark      &28.2 &0.0 & 3.8 &0.0       &17.4 & 3.6  &28.6 &11.6    \\
&\cmark   &\xmark    &7.7 &4.5 &0.0 &0.0       &0.0 &0.0  &0.0  &2.4  \\
&\xmark   &\cmark     &0.0 &18.2 &0.0 &0.0       &0.0 &0.0  &14.3  &3.0  \\
&\xmark   &\xmark  &64.1 &77.3 &96.2 &100.0       &82.6 &96.4  &57.1  &82.9    \\

\bottomrule
\end{tabular}
}
\label{tab:reasoning_correctness}
\end{table*}

 Table~\ref{tab:reasoning_correctness} presents the performance of two LLMs in understanding defect patterns by evaluating both the correctness of their predicted patterns and the accompanying reasoning. Results are reported for each defect pattern (e.g., RD, VR) along with the overall average of percentage. In this table, \cmark\ and \xmark\ denote correct and incorrect results, respectively.

\phead{Comparison of LLMs.}
Overall, \texttt{DeepSeek-R1 \& Direct+K} considerably outperforms \texttt{Qwen2.5-72b \& Direct} in both defect pattern detection and reasoning correctness. \texttt{Deepseek-R1} achieves an average ``\underline{P}attern \underline{C}orrect \& \underline{R}eason \underline{C}orrect'' (PCRC) rate of 45.7\%, whereas \texttt{Qwen2.5-72B} attains 11.6\% average PCRC. This indicates a substantial difference in their ability to both correctly identify defect patterns and justify their predictions. The majority of instances for \texttt{Deepseek-R1} fall into ``\underline{P}attern \underline{I}ncorrect \& \underline{R}eason \underline{I}ncorrect'' (PIRI) with an average of 53.0\%, while \texttt{Qwen2.5-72B} has a much higher average PIRI rate of 82.9\%. This is reasonable as incorrect reasoning typically leads to incorrect final predictions.

\phead{Reasoning Results by Defect Pattern.}
\texttt{DeepSeek-R1 \& Direct+K} achieves strong PCRC results for specific defect patterns, with 69.6\% on VR and 65.2\% on SS, but performs poorly on LV (11.5\%). In comparison, \texttt{Qwen2.5-72B \& Direct} has its best PCRC on PF (28.6\%) and RD (28.2\%), but fails completely on VR and SM (0.0\% PCRC), with SM showing a 100.0\% PIRI. Both models struggle with LV, where \texttt{Qwen2.5-72B} reaches 3.8\% PCRC.

\phead{Observations on Reasoning Accuracy and Errors.}
\texttt{DeepSeek-R1} demonstrates strong reasoning capability, with an average PCRC of 45.7\% and an average ``\underline{P}attern \underline{C}orrect \& \underline{R}eason \underline{I}ncorrect'' (PCRI) of 1.2\%. Notably, \texttt{Deepseek-R1} shows no instances (0.0\% average) of ``\underline{P}attern \underline{I}ncorrect \& \underline{R}eason \underline{C}orrect'' (PIRC). In contrast, \texttt{Qwen2.5-72B \& Direct} shows lower reasoning accuracy at 82.9\%, with a slightly higher average PCRI of 2.4\%. This model also has a 3.0\% average PIRC rate, with higher occurrences on VR (18.2\%) and PF (14.3\%) defect patterns.

\phead{Case Study.}
We present four illustrative examples corresponding to the correctness in defect pattern detection and the associated reasoning provided by LLMs.

\ulphead{Pattern Correct \& Reason Correct.}
Issue YARN-8907~\cite{YARN-8907} exemplifies a semantic inconsistent defect where a logging code at the end of a test case incorrectly prints \texttt{START}, misleadingly signaling the beginning of a test that has already completed. A developer has fixed this by changing \texttt{START} to \texttt{END}. The LLM successfully detects this semantics inconsistent issue in the logging code, demonstrating its ability to identify logical contradictions by aligning the natural language of the logging code with the behavioral context of the surrounding code.
\begin{lstlisting}[
    language=Java,
    label=list:LV,
    captionpos=t,
    breaklines=true,
    basicstyle=\ttfamily\footnotesize,
    tabsize=2,              
    breakatwhitespace=true, 
    breakautoindent=true,    
    breakindent=10pt,                                                              
    keepspaces=true,       
    columns=flexible,     
    showstringspaces=false 
]
// YARN-8907
LOG.info("--- START: testNotAssignMultiple ---");
\end{lstlisting}

\ulphead{Pattern Correct \& Reason Incorrect.}
The code in HDFS-15197~\cite{HDFS-15197} illustrates a case where the LLM identifies the correct defect pattern but for the wrong reason. Specifically, the model correctly flags the use of the \texttt{INFO} level for logging an exception as a potential issue related to the logging level. It recommends raising the level to \texttt{WARN} or \texttt{ERROR} to improve visibility. However, the developer's actual fix lowers the logging level to \texttt{DEBUG}, citing concerns over excessive log volume in scenarios involving frequent \texttt{ObserverRetryOnActiveException}. This case highlights a subtle challenge: although the LLM's reasoning appears intuitively valid, as many developers might assume that exceptions should be logged at higher severity levels, it fails to align with the intended logging behavior and context understood by the developer. This discrepancy might arise from the LLM's lack of project-specific background knowledge and familiarity with the system's established logging practices.

\begin{lstlisting}[
    language=Java,
    label=list:LV,
    captionpos=t,
    breaklines=true,
    basicstyle=\ttfamily\footnotesize,
    tabsize=2,              
    breakatwhitespace=true,  
    breakautoindent=true,    
    breakindent=10pt,                                                          
    keepspaces=true,      
    columns=flexible,     
    showstringspaces=false 
]
// HDFS-15197
LOG.info("Encountered ObserverRetryOnActiveException from {}." + " Retry active namenode directly.", current.proxyInfo);
\end{lstlisting}

\ulphead{Pattern Incorrect \& Reason Correct.}
This case highlights the result that correctly explains the reason for a defect but misclassifies its defect pattern. In the example from HDFS-14407~\cite{HDFS-14407}, the LLM accurately detects that improper string concatenation causes the \{\} placeholder to remain unsubstituted in the final log message. However, it labels the issue as a general string concatenation problem, instead of \textbf{VR-2} \textit{Placeholder–value mismatch}, which refers to mismatches between placeholders and variables that lead to confusing or misleading log output. In the actual fix, the developer removes the string concatenation and correctly supplies the variable as an argument. This case illustrates a key limitation: without domain-specific knowledge, an LLM’s reasoning may not align with the defect categorizations used by developers and may struggle to generate precise fix suggestions.

\begin{lstlisting}[
    language=Java,
    label=list:LV,
    captionpos=t,
    breaklines=true,
    basicstyle=\ttfamily\footnotesize,
    tabsize=2,               
    breakatwhitespace=true,  
    breakautoindent=true,    
    breakindent=10pt,                                                                
    keepspaces=true,       
    columns=flexible,     
    showstringspaces=false 
]
// HDFS-14407
LOG.warn("checkAllVolumes timed out after {} ms + maxAllowedTimeForCheckMs);
\end{lstlisting}

\ulphead{Pattern Incorrect \& Reason Incorrect.}
This example~\cite{Camel-Commit-0c1a589} demonstrates a complete failure in the LLM’s analysis, where both the reasoning and the prediction are incorrect. The model classifies the logging code as correct and reports no issues. However, the code actually contains a \textbf{SM-3} \textit{Misused Variables in the Message} defect, where an incorrect variable is used in the logging code. Specifically, the logging references \texttt{tempTarget} instead of the intended target variable, \texttt{target}, which can mislead users. This issue is later fixed by a developer by replacing \texttt{tempTarget} with \texttt{target}. The LLM fails to detect the defect, likely due to its inability to accurately interpret variable usage within the surrounding code context.

\begin{lstlisting}[
    language=Java,
    label=list:LV,
    captionpos=t,
    breaklines=true,
    basicstyle=\ttfamily\footnotesize,
    tabsize=2,               
    breakatwhitespace=true,  
    breakautoindent=true,    
    breakindent=10pt,                                                               
    keepspaces=true,       
    columns=flexible,      
    showstringspaces=false 
]
//Camel Commit 0c1a589
log.trace("Deleting existing file: " + tempTarget);
\end{lstlisting}

\greybox{\textbf{RQ3 Summary: } 
We find that even when LLMs correctly identify logging code defects, their accompanying reasoning is not always accurate or consistent with the code context. While correct predictions often align with valid reasoning, there are notable cases where the reasoning is incomplete or incorrect (e.g., a PIRI of 53.0\% and 82.9\%).
}

\section{Discussion}
\label{sec:discussion}

\subsection{Implications}
We summarize the implications of our study, where P1 - P3 are targeted at practitioners, while R1 - R2 provide insights for LLM researchers.

\phead{Implication P1: Actionable Guidelines for better logging.}
Our findings provide practical takeaways for software practitioners aiming to improve their logging practices. Practitioners can leverage our logging code defect patterns and the detailed scenario knowledge as guidelines (i.e., the results of RQ1). These patterns and scenarios can serve as a checklist during code development and reviews, or be integrated into developer training materials, to proactively improve the quality logging code and reduce defects.

\phead{Implication P2: Importance of domain knowledge.} Our findings in RQ2 show that enriching prompts with detailed scenario knowledge of defect patterns (i.e., \texttt{+K}) considerably improves LLM performance in detecting and reasoning the logging code defects. 
Practitioners can leverage this insight by incorporating detailed domain knowledge, such as concrete defect scenarios, explanations, and examples into prompt design or LLM fine-tuning pipelines.

\phead{Implication P3: Using LLM-generated reasoning with caution.} 
The results of RQ3 show that LLMs may produce incorrect or misleading reasoning. When integrating LLMs into development workflows (e.g., for code review or log monitoring), it is important to verify both the detection results and their justifications before acting on them.

\phead{Implication R1: Better structural code understanding for LLMs.}
We find that incorporating inter-procedural information (i.e., \texttt{+I}), such as control flow and data flow, does not consistently improve LLM performance and even degrades it. This suggests that current LLMs, though proficient in processing natural language and source code, may lack the structural understanding required to analyze code semantics effectively. Prior studies have demonstrated the utility of static analysis in traditional defect detection, highlighting the importance of structural code information. Future LLM research may explore ways to enhance models’ ability to interpret and utilize code structure, for example, by pretraining on program analysis artifacts, designing structure-aware representations, or integrating symbolic reasoning components.

\phead{Implication R2: Strengthening the LLMs' reasoning quality.}
Our RQ3 analysis reveals that even when LLMs correctly detect defect patterns in logging code, their accompanying explanations may be incomplete, inaccurate, or inconsistent with the actual code context. Conversely, LLMs can also produce logically sound reasoning for incorrect predictions, which indicates a gap between surface-level plausibility and genuine understanding. This discrepancy highlights the need to strengthen the reasoning capabilities of LLMs. Future research may explore techniques that align prediction outcomes with verifiable resoning process, or post-hoc verification mechanisms that assess the internal consistency of LLM-generated results.

\subsection{Threats to Validity}

\phead{Internal Validity.}
The validity of our taxonomy may be threatened by the diversity of our data sources. To mitigate this risk, we triangulate our findings through a comprehensive analysis of academic literature, issue tracking systems, and commit histories.

Since the dataset construction and the result analysis in this study involves a manual study, its validity can be influenced by the knowledge and experiences of the participants. To mitigate potential biases in such process, two authors independently perform the labeling, achieving substantial agreement, with an overall Cohen's Kappa value of 0.88.

\phead{External Validity.}
Our study investigate a specific set of four LLMs. While these represent a mix of open-source and closed-source models with varying capabilities, their performance in detecting and reasoning about logging defects may not be representative of all existing or future LLMs.

Since the \dataset benchmark is constructed using data from open-source Java projects, the number of studied subjects and programming languages may pose a threat to the study’s validity. To mitigate this, careful consideration goes into the selection of subjects. These analyzed projects are well-known and have gained considerable attention from developers and researchers, based on the stars on GitHub and existing research papers \cite{chen_characterizing_2017, hassani_studying_2018}.

Although our analysis is based on large and widely-used Java systems, our findings on defect prevalence and LLM effectiveness may not generalize to projects from other domains or those written in different programming languages. Future studies could explore the generalizability of our results across these different contexts.

\section{Conclusion}
\label{sec:conclusion}

In this paper, we present our comprehensive study on logging code defects, which identifies seven distinct patterns and constructs the \dataset benchmark with developer-verified real-world instances. Our evaluation of LLMs on \dataset reveals a notable gap between model predictions and human understanding, highlighting the limitations of current LLMs in reasoning about logging code and system runtime behavior. These findings underscore the value of our benchmark in guiding future research. Moreover, our work provides actionable insights for developers to avoid common defect patterns and lays a foundation for advancing LLM-based reasoning and defect detection in software logging.

\bibliographystyle{IEEEtran}
\bibliography{main}

\end{document}